
\documentstyle{amsppt-n}
\magnification=1200
\hsize=15.5truecm
\vsize=22.5truecm
\font\citefont=cmbx9
\def\A{{\Cal A}}
\def\B{{\Cal B}}
\def\Cc{{\Cal C}}

\def\D{{\Cal D}}

\def\I{{\Cal I}}

\def\O{{\Cal O}}

\def\L{{\Cal L}}
\def\M{{\Cal M}}

\def\Zc{{\Cal Z}}
\def\Z{{\Bbb Z}}

\def\C{{\Bbb{C}}}

\def\b#1,#2{B_L^{#1\vert#2}}

\def\pd#1,#2{\dfrac{\partial#1}{\partial#2}}
\def\sh#1#2{\hbox{$\Cal #1  #2 $}}

\def\hom{\sh{H}om\,}
\def\rest #1,#2{{#1}_{\vert #2}}
\def\iso{\kern.35em{\raise3pt\hbox{$\sim$}\kern-1.1em\to}
         \kern.3em}
\def\longiso{\kern.7em{\raise3pt\hbox{$\sim$}\kern-1.5em
              \longrightarrow}\kern.3em}

\def\gcoor#1#2,#3#4{(#1^1,\allowmathbreak\dots,#1^{#2},\allowmathbreak
                    #3^1,\allowmathbreak\dots,#3^{#4})}

\def\proof{\noindent {\it Proof.} \ }
\def\dim{\operatorname{dim}}
\def\cite#1{[{\citefont #1}]}
\def\bigsearrow{\hbox{\bigsym \char '46}}

\font\bigsym=cmsy10 scaled\magstep3
\def\fd#1{R^{#1}\pi_\ast}
\def\intXY{\int_{X/Y}\,}

\def\pic{\operatorname{Pic}}

\def\rg{{\B_X/B_Y}}
\def\berc{\operatorname{Ber}_{\B_X/\B_Y}}
\def\berr{\operatorname{Ber}_{\A_X/\A_Y}}
\def\intb{\int_{\A_X/\A_Y}\,}
\def\nb{\hat{\nabla}}
\def\nbb{\hat{\bar\nabla}}
\def\cub{{\vrule width4pt height6pt depth1.5pt\hskip1pt}}
\noTagsNumbers
\nosubheadingnumbers
\TagsOnRight
\def\Nc{{\Cal N}}
\def\obligedskip{\vbox{\vskip4mm}}
\topmatter
\title
LINE BUNDLES OVER FAMILIES OF (SUPER) \\ RIEMANN SURFACES.
II: THE GRADED CASE \dag
\\  \\ (March 1991 --- revised  September 1991 and October 1992)
\endtitle
\author
U\. Bruzzo\ddag\ {\rm and}\ J.A\. Dom\'\i nguez P\'erez\P
\endauthor
\affil
\ddag\thinspace Dipartimento di Matematica, Universit\`a
di Genova, Italia \\ \P\thinspace Departamento de
Matem\'atica Pura y Aplicada, \\ Universidad de Salamanca, Espa\~na
\endaffil
\address{(\ddag) Dipartimento di Matematica, Universit\`a
di Genova, Via L. B. Alberti 4, 16132 Genova, Italy. E-Mail:
{\smc bruzzo\@matgen.ge.cnr.it}}
\address{(\P) Departamento de Matem\'atica Pura y Aplicada,  Universidad
de Salamanca, Plaza de la Merced 1-4, 37008 Salamanca,
Spain}
\abstract{A relative Picard theory in the  context of graded manifolds is
introduced.  A Berezinian
calculus and a theory of connections over SUSY-curves
are systematically developed, and used to prove  a Gauss-Bonnet theorem for
line bundles in that setting and to discuss the validity of a
flatness theorem.}
\thanks{\dag\ Research partly
supported by the joint  CNR-CSIC research project `Methods and applications of
differential geometry in mathematical phys\-ics,' by `Gruppo Nazionale per la
Fisica Matematica' of CNR, by the Italian Ministry for University and Research
through the research project `Metodi geometrici e probabilistici
in fisica matematica,'
and by the Spanish CICYT through the research project `Geometr\'{\i}a
de las teor\'{\i}as gauge.'}
\endtopmatter
\document
\heading Introduction \endheading
{}From a geometric point of view, the application of moduli space techniques
to   the computation of quantum scattering amplitudes  \`a\ la Polyakov
for superstrings requires the introduction of the notion of super
Riemann surface  (also called SUSY-curve)   \cite{13}.
These structures are most conveniently studied within the framework of
Berezin-Le\u\i tes-Kostant's graded manifolds; physics suggests to define a
SUSY-curve as a family of complex graded manifolds of relative  dimension
(1,1), additionally endowed with a  {\sl conformal
structure}.

In this paper we complete the work we started in \cite{5}, where we considered
some facts concerning line bundles over families of ordinary Riemann
surfaces.
Here we prove  a version of the Gauss-Bonnet theorem
suitable to the context of SUSY-curves, and discuss the validity of a
flatness theorem for SUSY-curves (cf\. \cite{2,6}).
With `flatness theorem' we refer to the
classical result   that a holomorphic line bundle
on a Riemann surface is flat if and only if its Chern class vanishes.
One should notice that the conformal
structure of a SUSY-curve, as opposed to families of (1,1) dimensional
complex graded manifolds without additional structures, provides
essential tools for dealing with these problems.

The contents of this paper are as follows. In Section 2 we present the
extension to the graded case
of some concepts pertaining families of manifolds;
this includes a relative graded
de Rham theory, the notions of a relative Berezinian
sheaf and fiberwise Berezin integration,  a relative graded Serre
duality, and a suitable relative Picard theory.
In Section 3, after recalling the definition of SUSY-curve, we show how on such
objects a Berezinian differential calculus can be developed; one can even
introduce a relative de Rham-Berezin theory.

The last Section is
devoted to the demonstration of the announced results; in particular, the
statement of the Gauss-Bonnet theorem requires to consider {\sl conformal
connections} over the SUSY-curve.
The transposition of the flatness theorem to the graded setting
requires some care. Indeed, under very general assumptions,
which are for instance satisfied by the moduli space of SUSY-curves,
the relative flatness of a relative line bundle is sufficient, but not
necessary, to ensure that it has vanishing relative Chern class.
In order that relative flatness is also a necessary condition
one needs stronger assumptions, as we shall discuss in detail.

Some of these notions and results already appeared in the paper \cite{6},
even though the treatment given there is less systematic and precise;
for instance, the transition from the absolute to the relative case
is not obtained simply by replacing the sheaf cohomology functors by
the higher direct image functors. The case of a family of graded manifolds
is intrinsically different from the case of a single graded manifold,
basically because there is no notion of `local splitness' which is
compatible with the family structure \cite{16,20}. Thus,
several concepts in complex geometry do not seemingly admit a straightforward
generalization to this setting, as for instance Grauert's cohomology
base change theorem.

\obligedskip\heading Graded families and line bundles\endheading
\subheading{Graded complex  families}
Let us start by stating some facts about graded manifolds.
We shall denote by $(X,\B_X)$ a complex analytic graded manifold
\cite{14}, and by $(X,\O_X)$ the underlying complex analytic
manifold. The structural epimorphism $\B_X\to\O_X$ induces
an exact functor from the category of $\B_X$-modules to the category
of $\O_X$-modules, which is realized by the tensor product
$\otimes_{\B_X}\O_X$, and will be denoted by a tilde,
i.e\. $\widetilde\M\equiv\M\otimes_{\B_X}\O_X$.

A graded $\B_X$-module $\M$ is said to be {\sl coherent} \cite{17,19}
if every $x\in X$ has a neighborhood $U$ such
that $\rest{\M},U$ is finitely generated, and the kernel of any
morphism $\rest{\B_X^{p\vert q}},U\to\rest{\M},U$ is locally finitely
generated. The sheaves $\B_X^{p\vert q}$ are themselves
coherent.
If $\pi\colon (X,\B_X)\to (Y,\B_Y)$ is a  proper morphism of
complex graded manifolds (which means that the underlying morphism
$X\to Y$ is proper), and $\M$ is a coherent sheaf of graded
$\B_X$-modules, then for all $k\geq 0$ the  sheaf  $\fd{k}\M$ (the $k$th
higher direct image  of $\M$) is a  coherent graded $\B_Y$-module.

We say that a morphism $\pi\colon (X,\B_X)\to (Y,\B_Y)$
is {\sl flat} if $\pi_\ast\B_X$ is a flat graded $\B_Y$-module
(then $\pi_\ast\B_X$ is locally free over $\B_Y$).

\proclaim{Definition}
A  family of complex analytic graded manifolds
is a proper and  flat submersion $\pi\colon (X,\B_X)\to (Y,\B_Y)$
such that the underlying  morphism $\pi\colon (X,\O_X)\allowmathbreak\to
(Y,\O_Y)$ is a family of complex manifolds.
\footnote{We recall from \cite{5}
that a family of complex manifolds is a proper and flat
submersion $\pi\colon (X,\O_X)\to (Y,\O_Y)$
whose fibers are universally connected.}
\endproclaim
This definition is tailored so as to include the moduli space of SUSY-curves
(cf. \cite{10}).
The structure sheaf $\B_{X_y}$ of the fiber over $y$ is given by
$\B_{X_y}= \rest{\B_X},{X_y}\otimes_{(\B_Y)_y} k(y)$, with
$k(y)=(\B_Y)_y/{\frak m}_y\simeq\C$, where ${\frak m}_y$
is the maximal ideal of $(\B_Y)_y$.

The {\sl relative dimension} of the family is the pair of integers
$(n,n') = \dim (X,\B_X) -\dim (Y,\B_Y)$, which is also the dimension
of each fiber.

The following {\sl vanishing theorem} holds.
  \proclaim{Proposition}
Let $\pi\colon (X,\B_X)\to (Y,\B_Y)$ be a  family of
complex graded manifolds, of relative dimension $(n,n')$.
Then $\fd{k}\B_X = 0$ for all $k>n$.\endproclaim
\proof Let $\Nc$ be the nilpotent ideal of $\B_Y$, and, for the
sake of brevity,
let us denote $\M_k=\fd{k}\B_X$. One can prove that
$\widetilde\M_k\simeq\fd{k}\O_X$ by
resolving $\B_X$ with an injective complex $\I^\bullet$ of
$\B_X$-modules and noting that $\widetilde\I^\bullet$
is a resolution of $\O_X$ by injective abelian groups, which
can be used to compute the sheaves $\fd{k}\O_X$ (cf\. \cite{7}
Theorem 2.4.1). Now, $\fd{k}\O_X=0$ for $k>n$ for the
vanishing theorem for the underlying ordinary complex family,
so that $\M_k/\Nc\M_k=0$; since $\B_Y$
is a sheaf of  local rings,
and $\M_k$ is finitely generated as it is coherent,
we may apply the graded Nakayama Lemma \cite{3} to obtain $\M_k=0$
for $k>n$.\qed\enddemo

One defines in the obvious way the sheaves of {\sl relative graded derivations}
and {\sl relative graded differentials}, denoted by $\sh Der\,(\rg)$ and
$\Omega_\rg^1$; for any fiber $(X_y,\B_{X_y})$,
the sheaves $\Omega^1_{\B_{X_y}}$ of graded differentials are obtained from
the sheaf of relative graded differentials as in the non-graded case,
by restricting and tensoring by $k(y)$.

A similar definition of {\sl family of real  graded manifolds}
can be given, and, as a matter of fact, any family of complex
graded manifolds has an underlying real family
$\pi\colon (X,\A_X)\to (Y,\A_Y)$ (for convenience, the sheaves
$\A_X$ and $\A_Y$ are considered to be the complexifications
of the structure sheaves of the relevant real graded manifolds).
We denote by $\Omega^k_{\A_X/A_Y}$ the $k$th exterior power
of the sheaf $\Omega^1_{\A_X/A_Y}$ of relative  smooth
graded differentials, and call {\sl relative k-forms}
its sections. One can introduce a {\sl relative graded
de Rham theory} as in the non-graded case \cite{5},
by considering  a {\sl relative graded differential}
$d_r\colon\Omega^{k-1}_{\A_X/A_Y} \to\Omega^k_{\A_X/A_Y}$, and {\sl
relative graded de Rham sheaves} $DR^k_{\A_X/A_Y}\equiv \pi_\ast
\Zc^k_{\A_X/A_Y}/d_r \pi_\ast\Omega^{k-1}_{\A_X/A_Y}$, where
$\Zc^k_{\A_X/A_Y}$ denotes the closed relative $k$-forms
(cf\. \cite{18}). One can prove for
these sheaves  results analogous to  the non-graded case: there is a sheaf
isomorphism $DR^k_{\A_X/A_Y}\simeq R^k\pi_\ast\pi^{-1}\A_Y$ (cf\. \cite{16}),
and the projection $p\colon\Omega^k_{\A_X}\to\Omega^k_{\A_X/A_Y}$
induces a commutative diagram of $\C$-modules

$$\CD \Gamma(X,\Zc^k_{\A_X}) @>>> H^k(X,\C) @>>> \Gamma(Y,\fd{k}\C)\\ @V p VV
@VV p V @VVV \\ \Gamma(Y,\pi_\ast\Zc^k_{\A_X/A_Y}) @>>>
\Gamma(Y,DR^k_{\A_X/A_Y})  @>\sim>> \Gamma(Y,\fd{k}\pi^{-1}\A_Y)
\endCD \quad.\tag 2.1
$$
\rem{Remark} \rm
The relative graded de Rham sheaves do not coincide
with the relative de Rham sheaves of the underlying  ordinary family, contrary
to the case of a single graded manifold, where the graded and ordinary de Rham
theories do coincide \cite{9}. This in some sense reminds of
what happens in the
case of supermanifolds modelled over Grassmann algebras, cf\. \cite{1,3}.
\endrem
A family of real graded manifolds is said to be {\sl orientable}
if the underlying family of differentiable manifolds is orientable;
in this sense, any family of real graded manifolds underlying a family
of complex analytic graded manifolds is canonically oriented.

\subheading{Fiberwise Berezin integration}
Let $\pi\colon (X,\B_X)\allowmathbreak\to$ $(Y,\B_Y)$ be a family of
complex analytic graded  manifolds of relative dimension $(n,n')$.
Let $\{w_i,z_j,\eta_k,\theta_l\}$ be a {\sl relative coordinate system},
that is, a set of sections of $\B_X$, with the $w_i,z_j$'s even
and the $\eta_k, \theta_l$'s odd,
such that $\{w_i,\eta_k\}$ are coordinates on $(Y,\B_Y)$ and
$\{z_j,\theta_l\}$ are coordinates on the fibers $(X_y,\B_{X_y})$.
One has bases $\{\partial/\partial z_j , \partial/\partial \theta_l\}$
of $\sh Der\,(\rg)$ and $\{dz_j, d\theta_l\}$ of $\Omega_\rg^1$
(these can be considered also as bases for $\sh Der\,{\B_{X_y}}$
and $\Omega^1_{\B_{X_y}}$).

\proclaim {Definition} The relative Berezinian sheaf of the family
$\pi\colon (X,\B_X)\to (Y,\B_Y)$ is the Berezinian sheaf $\berc$ associated
with the locally free $\B_X$-module\break $\Omega_\rg^1$.
\endproclaim
Up to a change of parity, one can describe $\berc$ as the
cohomology group $H(K^\bullet)$ of the Koszul complex
$K^\bullet=S^\bullet_{\B_X} (\Pi \Omega_\rg^1 \allowmathbreak\oplus
\allowmathbreak (\Omega_\rg^1)^\vee)$  \cite {14},
where $\Pi$ is the parity change functor, and $^\vee$  denotes the dual module;
in this sense one denotes by $[dz_1\dots dz_n\otimes$
$\partial/\partial \theta_1\dots\partial/\partial \theta_{n'}]$ the
local basis of $\berc$ corresponding to a relative coordinate system
$\{w_i,z_j,\allowmathbreak\eta_k,\theta_l\}$
(as a matter of fact, $\berc\simeq\Pi^n H(K^\bullet)$).
Thus, if
$\{w'_i,z'_j,\eta'_k,\theta'_l\}$ is another relative coordinate system,
one has
$$[dz'_1\dots dz'_n\otimes
\frac{\partial}{\partial\theta'_1}\dots\frac{\partial}{\partial\theta'_{n'}}]=
[dz_1\dots dz_n\otimes
\frac{\partial}{\partial\theta_1}\dots\frac{\partial}{\partial\theta_{n'}}]
\operatorname{Ber} J \,,\tag 2.2$$
where $J$ is the Jacobian matrix of the coordinate transformation,
and $\operatorname{Ber}$ is the Berezin determinant.

For an oriented family of real graded manifolds
$\pi\colon (X,\A_X)\to (Y,\A_Y)$ of relative dimension $(m,m')$,
the sections of the corresponding Berezinian sheaf
$\berr$ (defined as in the complex analytic case) can be interpreted
in a natural way as `graded volume forms' on the family,  and one has a {\sl
fiberwise Berezin
integration}
$$\intb\colon\pi_\ast\berr\to\A_Y\,,\tag 2.3$$
which is defined as follows (cf\. \cite{11}, and also \cite {4}
for the absolute case).
\proclaim{Definition} (1) Let $\{y_i,x_j,\mu_k,\nu_l\}$ be a
relative coordinate system, and let $\omega$ be a section of
$\pi_\ast\berr$ such that
$$\omega = [dx_1\dots dx_m\otimes\frac{\partial}{\partial\nu_1}
\dots\frac{\partial}{\partial\nu_{m'}}]\sum_{\alpha,\beta}
f_{\alpha\beta}(y,x)
\mu_\alpha\nu_\beta\,,$$
where the $f_{\alpha\beta}$'s are sections of $\Cc_X$,
$\alpha$, $\beta$ are multi-indices of the type
$\beta=\{\beta_1,\dots,\beta_s\}$ with $1\leq\beta_1<\dots<\beta_s\leq m'$,
and $\nu_\beta=\nu_{\beta_1}\wedge\dots\wedge\nu_{\beta_s}$.
Then one defines
$$\intb \omega = \sum_\alpha \bigl( \intXY f_{\alpha\varpi} dx_1\dots dx_m
\bigr)\,\mu_\alpha\,,$$
where $\varpi$ is the multi-index $\{1,2,\dots,m'\}$, and $\intXY$ is the
fiberwise integration {\rm \cite {5}} of sections of the sheaf
$\pi_\ast\Omega^m_{X/Y}$.

(2) For a generic section $\omega$, one defines $\intb$ by additivity,  using
a partition of unity and (1).
\endproclaim

\subheading{Relative Serre duality}
Let $\pi:(X,\B_X)\to(Y,\B_Y)$ be a family of complex graded
manifolds of relative dimension $(n,n')$. In this context,
the Berezinian sheaf $\berc$,  analogously to the sheaf
$\kappa_{X/Y}$ of relative  holomorphic $n$-forms in the
non-graded case, plays the role of {\sl dualizing sheaf\/};
that is, if $\M$ is a coherent $\B_X$-module, there is a
canonical isomorphism of $\B_Y$-modules \cite {15,16,20}
$$R\pi_\ast R\hom_{\B_X}(\M,\berc(-n))\simeq
R\hom_{\B_Y}(R\pi_\ast\M,\B_Y)\,.\tag 2.4$$

This assertion can be proved as in the
non-graded case \cite{8}, and one obtains
in particular (cf\. \cite {5})
$$R^n\pi_\ast\berc\simeq
(\pi_\ast\B_X)^\vee\,.\tag 2.5$$

\subheading{The Picard group of a graded family}
A {\sl line bundle} over a complex graded manifold $(X,\B_X)$
is a locally free $\B_X$-module either
of rank $(1,0)$ or $(0,1)$; the parity change functor $\Pi$ establishes
a one-to-one correspondence between the groups of isomorphism classes of
line bundles of fixed rank. The {\sl Picard group} of $(X,\B_X)$
is the group $\pic (X,\B_X)$ of isomorphism classes of
line bundles, disregarding parity.

On the analogy of the non-graded case, the elements in
$\pic (X,\B_X)$ are characterized by \v Cech 1-cocycles
of the sheaf $(\B_X)^\ast_0$ (invertible and even sections of $\B_X$),
called {\sl transition functions}, and there is an isomorphism
$\pic (X,\B_X)\simeq H^1(X,(\B_X)^\ast_0 )$.
If $(X,\O_X)$ is the complex analytic manifold underlying $(X,\B_X)$, the
structural projection $\B_X\to\O_X$ induces a morphism
$$\eqalign{\pic (X,\B_X) & \to\pic (X) \cr
\L & \mapsto \L\otimes_{\B_X} \O_X \equiv \widetilde\L\cr}\tag 2.6 $$
which has, in general, a kernel and a cokernel.
In the particular case that
$(\B_X)^\ast_0 \simeq \O^\ast_X$, (2.6) is of course an isomorphism
(for instance, this happens if the odd dimension of $(X, \B_X)$ is 1).

One can, however, identify the {\sl Chern classes}
$c_1(\L)=c_1(\widetilde\L)\in
H^2(X,\Z)$, where $c_1(\L)$ is defined as  minus the image of $\L\in\pic
(X,\B_X)$ via the cohomology morphism induced by the first line
of the commutative diagram
$$ \CD 0 @>>> \Z @>>> (\B_X)_0 @> {\exp 2\pi i} >> (\B_X)^\ast_0 @>>>0 \\
@. @| @VVV @VVV \\
0 @>>> \Z @>>> \O_X @> {\exp 2\pi i} >> \O_X^\ast @>>>0 \endCD
\quad;\tag 2.7$$
the desired identification is then provided by the cohomology diagram
induced by (2.7).

The concept of  Picard group and related notions  in the graded setting can be
extended  to the context of families, exactly as  it happens in the non-graded
case \cite {5}. The {\sl relative Picard sheaf} of the family $\pi\colon
(X,\B_X) \to (Y,\B_Y)$ is the sheaf $\fd{1}(\B_X)^\ast_0$; the {\sl
(restricted)
relative
Picard group} is the space $\pic (\B_X/\B_Y) = \Gamma(Y,\fd{1}(\B_X)^\ast_0)$,
and there is a functorial morphism $\phi\colon\pic (X,\B_X)\to
\pic (\B_X/\B_Y)$, $\L \mapsto [\L]$.
The {\sl relative Chern class} $c_1$ is minus the  morphism
$\fd{1}(\B_X)^\ast_0\to\fd{2}\Z$ induced by (2.7)
(if the underlying complex family is a family of Riemann surfaces,
the Chern class can be regarded as a section of $\fd{2}\C$); one denotes in the
same way the corresponding morphism between spaces of  global sections, and one
proves that $c_1$ is functorial and commutes with $\phi$  as in the non-graded
case.  A section of $\fd{1}(\B_X)^\ast_0$ is said to be {\sl flat}
if it lies in the image of the sheaf morphism
$\fd{1}\pi^{-1}(\B_Y)^\ast_0\to\fd{1}(\B_X)^\ast_0$
induced by the natural morphism $\pi^{-1}(\B_Y)^\ast_0\hookrightarrow
(\B_X)^\ast_0$.

\obligedskip\heading SUSY-curves\endheading
\subheading{Basic definitions}
Let us recall a brief motivation, in terms of conformal structures,
of why a `super Riemann surface' is not simply a (1,1) dimensional
complex graded manifold.
In the ordinary case,
given a two-dimensional manifold $X$, the complex structures on $X$
are in a one-to-one correspondence with the conformal classes of
Riemannian structures on $X$, and conformal changes of the metric structure
correspond to holomorphic maps, i.e\. they map the vector field
$\partial/\partial z$ into a multiple of itself.
One might wonder what is the analogue of this situation
in the graded case. If $(X,\B_X)$ is a complex graded manifold
of dimension (1,1), with local coordinates $(z,\theta)$,
locally the  even vector field $\partial/\partial z$ has an odd `square root':
if $D=\partial/\partial \theta+\theta\partial/\partial z$,
then $D^2=\partial/\partial z$. We say  that $(X,\B_X)$ is endowed
with a conformal structure if the local $D$'s define a global rank (0,1)
submodule of the tangent sheaf (one can equivalently
require the existence of a maximally nonintegrable rank (0,1) submodule
of the tangent sheaf, and then proves that that submodule is locally
generated by $D$).
Thus, in the graded case a conformal structure is richer
than a complex structure, and it comes out that in order to extend
to the graded setting several classical constructions one needs indeed
a conformal structure.

\proclaim{Definition} A graded Riemann surface, or SUSY-curve
{\rm \cite {13}},
is a family of complex graded manifolds $\pi\colon (X,\B_X)\to (Y,\B_Y)$
of relative dimension $(1,1)$, endowed with a
locally free rank (0,1) subsheaf $\D$ of $\sh Der\,(\rg)$,
such that the morphism
$$\CD \D\otimes_{\B_X}\D @>[\ ,\ ]\ \text{mod}\,\D>>
\sh Der\,(\rg)/\D\endCD\tag 3.1$$
is an isomorphism of graded $\B_X$-modules.
\endproclaim
Here $[\ ,\ ]$ is the graded Lie bracket. The subsheaf $\D$ is called
a {\sl conformal structure\/};
we shall briefly denote a SUSY-curve by $(X/Y,\D)$.
For any $y\in Y$, the subsheaf $\D_y = \rest{\D},{X_y}\otimes_{(\B_Y)_y}k(y)$
of $\sh Der\, \B_{X_y}$ endows the
fiber $(X_y,\B_{X_y})$ of a SUSY-curve $(X/Y,\D)$
over $y$ with a structure
of SUSY-curve (as a family over the point $y$),
which we denote briefly by $(X_y,\D_y)$.

We would like to recall some results about SUSY-curves,
which relate Definition 3.1 to the more intuitive notions of SUSY-curve
as suggested by analogy with the non-graded case or by physical motivations
(proofs can be found in   \cite{10,12,13}).
\proclaim {Lemma} Any SUSY-curve $(X/Y,\D)$ admits  relative coordinate
systems $\{w_i,z,\allowmathbreak\eta_k,\theta\}$, called conformal, such that
$\D$ is locally generated by the odd derivation
$$D={\partial\over\partial\theta}+\theta{\partial\over\partial z}\,.\tag 3.2$$
\qed\endproclaim

We shall denote conformal coordinates by $(z,\theta)$,
and call {\sl conformal} the coordinate changes which preserve
conformal coordinates. A simple computation shows that conformal
changes have the form
$$z'  =\varphi(z) \pm \theta\psi(z)\sqrt
{ \frac{\partial\varphi}{\partial z} + \psi(z) \frac{\partial\psi}
{\partial z}}
=\varphi(z) \pm \theta\psi(z)\sqrt
{ \frac{\partial\varphi}{\partial z}}
\,,$$
$$
\theta'=\psi(z) \pm \theta\sqrt
{ \frac{\partial\varphi}{\partial z} + \psi(z) \frac{\partial\psi}
{\partial z}}
$$
where $\varphi$ (resp\. $\psi$) is an even (resp\. odd)
holomorphic graded  function which additionally depends
on the coordinates of $(Y,\B_Y)$.

If the `parameter space' $(Y,\B_Y)$ has no odd parameters (i.e\. it is an
ordinary manifold), then the SUSY-curve is {\sl split\/}, namely, the structure
sheaf of the total space is an exterior algebra.
\proclaim{Lemma} Let $\pi\colon (X,\B_X)\to (Y,\O_Y)$ be
a family of complex analytic graded manifolds of relative dimension $(1,1)$,
with $(Y,\O_Y)$ an ordinary complex manifold. Then: \roster
\item $\B_X \simeq \Lambda^\bullet_{\O_X} L$, where $L$ is a line bundle
on $(X,\O_X)$;
\item the specification of a conformal structure $\D$ on the family
is equivalent to $L$ being
a ``relative spin structure'' on $(X,\O_X)$, that is,
to an isomorphism
$ L \otimes_{\O_X} L \simeq \kappa_{X/Y}$,
where $\kappa_{X/Y}$ is the sheaf of  relative holomorphic 1-differentials
on $(X,\O_X)$; moreover, one has $\D^\ast\simeq L\otimes_{\O_X}\Pi\B_X$.
\endroster
\qed\endproclaim

Notice that the particular case of a SUSY-curve over a point
(for instance, the fiber of a generic SUSY-curve)
satisfies the hypotheses of this Lemma.

\proclaim{Proposition} If $(X/Y,\D)$ is a SUSY-curve, there is a canonical
isomorphism of $\B_X$-modules
$$\D^\vee \longiso \berc\,.\tag 3.3$$
\endproclaim
\proof In conformal coordinates $(z,\theta)$, the module $\D^\vee$ has
a local basis $[d\theta]$ given by the image of $d\theta\in\Omega_\rg^1$
under the epimorphism $\Omega_\rg^1\to\D^\vee$ (dual of the natural
inclusion), and $\berc$ has a basis $[dz\otimes\partial/\partial \theta]$.
One defines the isomorphism (3.3) locally by letting
$[d\theta]\mapsto[dz\otimes\partial/\partial \theta]$,
and proves that it does not depend on the choice of the conformal coordinates.
\qed\enddemo

\subheading{Berezinian calculus}
If $\pi\colon (X,\B_X)\to (Y,\B_Y)$ is a family of complex graded manifolds
of relative dimension $(1,1)$, one can define, analogously to the non-graded
case, the {\sl sheaves of smooth relative $(p,q)$-forms}
$$\Omega^{p,q} = (\A_X \otimes_{\B_X} (\Omega_\rg^1)^p)
\otimes_{\bar\B_X} (\bar\Omega_\rg^1)^q\,,\qquad p,q=0,1,\tag3.4$$
where $(X,\A_X)$ is the graded smooth manifold underlying
$(X,\B_X)$, and a bar denotes  complex conjugation.
Moreover,
$\Omega^{1,0}\oplus\Omega^{0,1}\simeq \Omega_{\A_X/\A_Y}$, and
the graded relative differential induces differential operators
$$\partial_r\colon\A_X\to\Omega^{1,0} \,,\qquad
\bar \partial_r\colon\A_X\to\Omega^{0,1} \,.\tag 3.5$$
However,  relative $(p,q)$-forms   are not the correct forms
to construct a differential calculus, because there are no top-degree forms,
and
they are not related to  `fiberwise integrable objects'. Therefore, it is
convenient to define a new kind of forms in terms of Berezinian sheaves;
such forms are naturally defined, and exist only, on SUSY-curves.
\proclaim {Definition}  Let $(X/Y,\D)$ be a SUSY-curve,
$d_r\colon\B_X\to\Omega_\rg^1$ the (graded) relative differential,
and $\varrho\colon\Omega_\rg^1\to\D^\vee
\simeq\berc$ the natural epimorphism. The holomorphic Berezin differential
$\hat\partial$ is the  $\pi^{-1}\B_Y$-linear morphism
$$\hat\partial = \varrho\circ d_r\colon\B_X\to\berc\,.\tag 3.6$$
\endproclaim
In a conformal coordinate system $(z,\theta)$  one has $\hat\partial f=
[dz\otimes\partial/\partial\theta] D(f)$, with $D$ as in (3.2).
\proclaim {Lemma} There is an exact sequence of $\pi^{-1}\B_Y$-modules
$$0@>>> \pi^{-1}\B_Y @>>> \B_X @>\hat\partial>> \berc @>>>0\,.\tag 3.7$$
\endproclaim
\proof It is enough to prove  exactness on the stalks, which
permits us to use conformal coordinates $(z,\theta)$.
Trivially, a section $f$ in $\pi^{-1}\B_Y$ satisfies $\hat\partial f=0$;
furthemore,
$0=\hat\partial f = [dz\otimes\partial/\partial\theta] D(f)$,  with
$f=f_0(z)+\theta f_1(z)$, implies that $\partial f_0/\partial z  = 0$ and $f_1
=0$, so that $f$ is a section of $\pi^{-1}\B_Y$. To conclude, one verifies that
the section  $f=\int g_1(z) dz +\theta g_0(z)$ of $\B_X$ is a
pre-image of the section
$[dz\otimes\partial/\partial\theta] (g_0(z) + \theta g_1(z))$ of $\berc$ under
$\hat\partial$
(where $\int dz$ is the indefinite path integral).
\qed\enddemo
Since one also has the exact sequence of $\pi^{-1}\B_Y$-modules
$0@>>> \pi^{-1}\B_Y @>>> \B_X @>d_r>> \Zc^1_{\B_X/\B_Y} @>>>0$,
one can identify the sheaves
$\berc$ and $\Zc^1_{\B_X/\B_Y}$ as $\pi^{-1}\B_Y$-modules;
however, this isomorphism
shall be regarded in a sense as accidental, and will not be exploited
in the sequel.

\proclaim{Definition} We define the sheaves of Berezin $(p,q)$-forms as the
sheaves over $X$
$$\A^{p,q} = (\A_X \otimes_{\B_X} (\berc)^p)\otimes_{\bar\B_X}
(\overline{\operatorname{Ber}}_{\B_X/\B_Y})^q\,,\qquad p,q=0,1\,.\tag 3.8$$
\endproclaim
Sections of these sheaves have been elsewhere called
$(p/2,q/2)$-differentials  \cite {6}.
\proclaim{Lemma} There is a canonical isomorphism of $\A_X$-modules
$$\A^{1,1}=\A^{1,0}\otimes_{\A_X}\A^{0,1} \longiso \berr\,.$$
\endproclaim
\proof In conformal coordinates $(z,\theta)$ the $\A_X$-module $\A^{1,1}$
has a basis $[dz\otimes\partial/\partial \theta] \otimes
[d\bar z\otimes\partial/\partial \bar\theta]$, and
$\berr$ has a basis $[dz \,d\bar z\otimes\partial/\partial \theta\,
\partial/\partial\bar\theta]$. One defines locally this isomorphism
 by letting
$[dz\otimes\partial/\partial \theta] \otimes
[d\bar z\otimes\partial/\partial \bar\theta] \mapsto
[dz\, d\bar z\otimes\partial/\partial \theta \,\partial/\partial\bar\theta]$,
and then proves that it is independent of the choice of conformal coordinates.
\qed\enddemo
The epimorphism $\varrho\colon\Omega_\rg^1\to\berc$ induces epimorphisms
$\varrho^{p,q}\colon\Omega^{p,q}\to\A^{p,q}$;
composing with the differential operators $\partial_r$
and $\bar \partial_r$ (cf\. (3.5)) one has:
\proclaim {Definition} The morphisms
$$\delta = \varrho^{1,0}\circ \partial_r\colon\A_X\to\A^{1,0}\,,\qquad
\bar\delta = \varrho^{0,1}\circ\bar \partial_r\colon\A_X\to\A^{0,1}$$
are  the smooth Berezin   differentials.
\endproclaim
The differential $\delta$ is $\bar\B_X$-linear, while
$\bar\delta$ is $\B_X$-linear.
In conformal coordinates $(z,\theta)$ the expression of the
smooth Berezin differentials is
$$\delta f=[dz\otimes\partial/\partial\theta] D(f) \,,\qquad
\bar\delta f=[d\bar z\otimes\partial/\partial\bar\theta] \bar D(f) \,.
$$

\proclaim {Proposition} There are exact sequences, respectively
of $\B_X$-modules and $\bar\B_X$-modules,
$$0@>>> \B_X @>>> \A_X @>\bar\delta>> \A^{0,1} @>>>0 \,,\qquad
0@>>> \bar\B_X @>>> \A_X @>\delta>> \A^{1,0} @>>>0 \,.$$
\endproclaim
\proof One proves the exactness of the first sequence (the second is
analogous) in the stalks, which allows us to use
conformal coordinates $(z,\theta)$.
If $f$ is a section in $\B_X$, in the induced coordinates
$(z,\bar z,\theta,\bar\theta)$ in $\A_X$ it does not depend on
 $\bar z$ and $\bar\theta$, so that $\bar\delta f=0$;
moreover, $0=\bar\delta f = [d\bar z\otimes\partial/\partial\bar\theta] \bar
D(f)$, with $f=f_0(z,\bar z)+\theta f_1(z,\bar z) + \bar\theta f_2(z,\bar z)
+\theta\bar\theta f_3(z,\bar z)$, implies
$\partial f_0/\partial \bar z = \partial f_2/\partial \bar z = 0$
and $f_2=f_3 =0$, so that $f$ is a section of $\B_X$.
Finally, $\bar\delta$ is surjective because it is the composition of
surjective morphisms.
\qed\enddemo
We extend the smooth Berezin differentials to morphisms
$\delta\colon\A^{0,1}\to\A^{1,1}$
and $\bar\delta\colon\A^{1,0}\to\A^{1,1}$
by letting $\delta(f\otimes\bar\omega)=\delta f\otimes\bar\omega$ and
$\bar\delta(f\otimes\omega)=\bar\delta f\otimes\omega$.
In conformal coordinates $(z,\theta)$, this implies
$\delta([d\bar z\otimes\partial/\partial\bar\theta] g) =
- [dz d\bar z\otimes\partial/\partial\theta \partial/\partial\bar\theta]
D(g)$, and $\bar\delta([dz\otimes\partial/\partial\theta] g) =
- [dz d\bar z\otimes\partial/\partial\theta \partial/\partial\bar\theta]
\bar D(g)$.

Thus, the smooth Berezin differentials act as sheaf morphisms
$$\eqalign{&\delta\colon\A^{p,q}\to\A^{p+1,q}\,,\quad p=0,\,q=0,1\,,\cr
&\bar\delta\colon\A^{p,q}\to\A^{p,q+1}\,,\quad p=0,1,\,q=0\,.\cr}$$

\proclaim{Lemma} The identity
$\delta\circ\bar\delta + \bar\delta\circ\delta = 0$ holds.
\endproclaim
\proof The claim reduces to proving the easy identity
$D\circ\bar D +\bar D\circ D = 0$.
\qed\enddemo

\proclaim {Proposition} The following  sequences,
respectively of $\B_X$-modules and $\bar\B_X$-modules, are exact.
$$\eqalign{&0@>>> \berc @>>> \A^{1,0} @>\bar\delta>> \A^{1,1} @>>>0 \,,\cr
&0@>>>\overline{\operatorname{Ber}}_{\B_X/\B_Y}
                    @>>> \A^{0,1} @>\delta>> \A^{1,1} @>>>0 \,.\cr}
$$
\endproclaim
\proof One tensors the exact sequences in Proposition 3.10 by
$\otimes_{\bar\B_X}\overline{\operatorname{Ber}}_{\B_X/\B_Y}$
(or $\otimes_{\B_X} \berc$).
\qed\enddemo

\proclaim {Definition} The (relative) de Rham-Berezin sheaf $DRB_{X/Y}$
of the SUSY-curve $(X/Y,\D)$ is the sheaf over $Y$
$$DRB_{X/Y}\equiv\frac{\pi_\ast\A^{1,1}}{\bar\delta\pi_\ast\A^{1,0}
\oplus\delta\pi_\ast\A^{0,1}}\,.$$
\endproclaim
The relationship between the relative de Rham sheaf and the
de Rham-Berezin sheaf can be established by constructing a morphism
$$\Psi\colon\Zc^2_{\A_X/\A_Y}\to\A^{1,1}\tag 3.9$$
as follows: a section $\xi$ of $\Zc^2_{\A_X/\A_Y}$ is locally
$\xi=d_r\omega$ for a section $\omega$ of $\Omega^1_{\A_X/\A_Y}$
(Poincar\'e lemma); then, if $\omega=\omega^{1,0}+\omega^{0,1}$ with
$\omega^{p,q}$ section of $\Omega^{p,q}$, one defines locally $\Psi$
by the formula
$$\Psi (\xi)=\bar\delta (\varrho^{1,0}(\omega^{1,0})) +
              \delta (\varrho^{0,1}(\omega^{0,1}))\,.$$
This definition is in fact global, because if $\xi=d_r\omega'$,
with $\omega'=\omega + d_r f$ and $f$ a section of $\A_X$, one has
$\bar\delta (\varrho^{1,0}((d_r f)^{1,0})) +
              \delta (\varrho^{0,1}((d_r f)^{0,1})) = 0$.

Composing the morphism $\pi_\ast\Zc^2_{\A_X/\A_Y}\to\pi_\ast\A^{1,1}$
induced by (3.9) with the projection onto the quotient $DRB_{X/Y}$, one
obtains a sheaf morphism $\pi_\ast\Zc^2_{\A_X/\A_Y}\to DRB_{X/Y}$;
an easy computation shows that it factorizes through $DR^2_{\A_X/\A_Y}$;
taking global sections, one has a commutative diagram of $\C$-modules
$$\CD
\Gamma(X,\Zc^2_{\A_X/\A_Y})@>>>\Gamma(X,\A^{1,1})\\
@V p VV @VVV \\
\Gamma(Y, DR^2_{\A_X/\A_Y}) @>>> \Gamma(Y,DRB_{X/Y})
\endCD \quad.\tag 3.10
$$
\proclaim{Proposition} {\rm (Stokes theorem)} Let $\tau$ and $\bar\tau'$ be
sections of $\pi_\ast\A^{1,0}$ and $\pi_\ast\A^{0,1}$ respectively. Then
$$\intb \bar\delta \tau = \intb \delta \bar\tau' = 0 \,.$$
Thus, fiberwise Berezin integration induces a morphism
$$\intb\colon DRB_{X/Y}\to\A_Y \,.$$
\endproclaim
\proof It is enough to prove this result when, after choosing
conformal coordinates $(z,\theta)$, one has
$\tau=[dz\otimes\partial/\partial\theta] f$
and $\bar\tau'=[d\bar z\otimes\partial/\partial\bar\theta] g$,
with $f,g$ sections of $\A_X$; then
$\intb \bar\delta \tau = \int_{X/Y} \bar D (f)_{\theta\bar\theta}
dz d\bar z$ and $\intb \delta \bar\tau' = \int_{X/Y}
D(g)_{\theta\bar\theta} dz d\bar z$.
One verifies that $\bar D (f)_{\theta\bar\theta}\allowmathbreak dz\wedge
d\bar z$ and $D (g)_{\theta\bar\theta} dz\wedge d\bar z$
are ordinary relative 2-forms, exact with respect to
$d_r$, and one proves the first assertion from
Stokes' theorem for $\int_{X/Y}$.
This yields a morphism from the quotient presheaf
$\pi_\ast\A^{1,1}/(\bar\delta\pi_\ast\A^{1,0}
\oplus\delta\pi_\ast\A^{0,1})$ into $\A_Y$,
and one concludes by factorizing through the associated sheaf $DRB_{X/Y}$.
\qed\enddemo

\obligedskip
\heading Line bundles over SUSY-curves\endheading
\subheading{Flatness theorem}
Let $(X/Y,\D)$ a SUSY-curve. In order to derive a characterization of
the flat sections of the relative Picard sheaf $\fd{1}(\B_X)^\ast_0$,
which parallels the analogous result in the non-graded case, let us
consider the commutative diagram
$$ 
\CD
@.      @.    0     \\
@.     @.    @AAA            \\
@.      @.       \berc   \\
@.     @.    @AAA            \\
0  @>>> \Z @>>>  \B_X  @>>> \B_X^\ast @>>> 0 \\
@.     @|     @AAA       @AAA            \\
0  @>>> \Z @>>>  \pi^{-1}\B_Y  @>>> \pi^{-1}\B_Y^\ast @>>> 0 \\
\endCD\quad.\tag 4.1$$
Applying the higher direct image functor, one has $R^2\pi^\ast\B_X =0$,
so that from the central vertical sequence one obtains an epimorphism
$$\alpha\colon R^1\pi_\ast\berc\to R^2\pi_\ast\pi^{-1}\B_Y \,,\tag 4.2$$
and from the bottom horizontal sequence a morphism
$$\beta\colon R^2\pi_\ast\Z \to R^2\pi_\ast\pi^{-1}\B_Y \,.$$
\proclaim{Lemma} The morphism
$\beta$ is injective.\endproclaim
\proof After taking derived functors in the commutative diagram
$$\CD \Z @= \Z \\ @VVV @VVV \\ \pi^{-1}\B_Y @>>> \pi^{-1}\O_Y \endCD$$
the claim follows from the analogous  result in the non-graded case (\cite{5}
Lemma 3.10).  \qed\enddemo
{}From (4.1) one has therefore the commutative diagram
$$ 
\CD
0 \\
@AAA \\
R^2\pi_\ast\pi^{-1}\B_Y \\
@A \alpha AA \\
R^1\pi_\ast\berc \\
@AAA \\
R^1\pi_\ast \B_X        @>>> R^1\pi_\ast \B_X^\ast        @> \dsize {c_1} >>
               R^2\pi_\ast\Z    @>>>      0   \\
@AAA @AAA\\
R^1\pi_\ast\pi^{-1}\B_Y @>>> R^1\pi_\ast\pi^{-1}\B_Y^\ast @>>> 0  \\
\endCD\quad.$$
Restricting this diagram to the even parts of the modules, one can deduce by a
simple diagram chasing that any flat section of the relative Picard sheaf
$\fd{1}(\B_X)^\ast_0$ has a vanishing relative Chern class.

In order to prove the converse, one should demonstrate that the epimorphism
$\alpha_0\colon R^1\pi_\ast(\berc)_0 \to R^2\pi_\ast\pi^{-1}(\B_Y)_0$
is bijective.
However, this  is false in general;  it is possible for instance to
construct SUSY-curves having the following properties:

(i) the base space $Y$ is a point, so that the derived functors reduce to
cohomology groups;

(ii) $\dim H^1(X,(\berc)_0)=2$, while $\dim H^2(X,\pi^{-1}(\B_Y)_0)=1$.

\noindent Such  an example  can be found in \cite{6}, page 294.

A condition assuring that $\alpha$ is an isomorphism is the analogue of
the universal connectedness that we have assumed in the ordinary case, namely,
$\pi_\ast\B_X\simeq\B_Y$.\footnote{One should notice that the moduli space of
SUSY-curves does not fulfill such a requirement.}
The proof that under this
condition $\alpha$ is in fact an isomorphism goes as in the ordinary case (cf\.
\cite{5} Lemma 3.9).

Thus, we arrive at the following result.

\proclaim{Theorem} Let $(X/Y,\D)$ be a SUSY-curve, and assume that
$\pi_\ast\B_X\simeq\B_Y$. Then,
any $y\in Y$ has an open neighborhood $V$ such that  a   section of
$\fd{1}(\B_X)^\ast_0$ has  a vanishing  relative Chern class if and
only if it is  flat. \qed\endproclaim

\rem{Remark} \rm The fact that the morphism $\alpha_0$
is not bijective implies that an analogue of Grauert's
cohomology base change theorem does not hold for SUSY-curves.
Indeed, if such a theorem held, one should have
$$\fd1\berc\otimes_{\B_Y}k(y)\simeq
H^1(X_y,\operatorname{Ber}_{\B_{X_y}})\,.$$
Since
$\operatorname{Ber}_{\B_{X_y}}\simeq\kappa_{X_y}\oplus\kappa_{X_y}^{1/2}$, with
$\kappa_{X_y}^{1/2}$ a  spin structure over the fiber $X_y$,  one has
$\fd1\berc\otimes_{\B_Y}k(y)\simeq\C\oplus \C^q$, where either $q=0$ or $q=1$
for a generic position in the moduli space. On the other hand,
$\fd2\pi^{-1}\B_Y\otimes_{\B_Y}k(y)\simeq
(\B_Y)_y\otimes_{(\B_Y)_y}k(y)\simeq\C$.
Restricting to even parts, Nakayama Lemma would imply that $\alpha$
is bijective.
\endrem

\subheading{Conformal connections and Gauss-Bonnet theorem}
Let $(X,\B_X)$ be a complex graded manifold, and $\L$ a line bundle over it.
As in the ordinary case, one can prove that $c_1 (\L)=\frac{i}{2\pi} [K]$,
where $[K]$ is the de Rham cohomology class of a (smooth) curvature
form $K$ on $\L$. A procedure analogous to that employed in \cite {5} for
families of complex manifolds allows us to define, for a family of
complex analytic graded manifolds $\pi\colon(X,\B_X)\to(Y,\B_Y)$, a
{\sl (smooth)
relative graded connection} over $\L$ as a morphism of $\C$-modules
$\nabla_r\colon \L\to \Omega^1_{\A_X/\A_Y}\otimes_{\B_X}\L$
which satisfies the Leibniz rule $\nabla_r (f\sigma)=
d_rf\otimes\sigma + f\nabla_r (\sigma)$. The {\sl curvature $K_r$} of
a relative graded connection $\nabla_r$ is the $\C$-module morphism
$K_r=\nabla_r^2\colon \L\to \Omega^2_{\A_X/\A_Y} \otimes_{\B_X}\L$, which
is $\B_X$-linear and determines a global section $K_r$ of
$\Omega^2_{\A_X/\A_Y}$; this is closed under the relative differential,
that is, $K_r \in \Gamma (X,\Zc^2_{\A_X/\A_Y})$, and its projection $[K_r]
\in \Gamma (Y,DR^2_{\A_X/\A_Y})$ does not depend on the relative
connection over $\L$. If $[\L]$ denotes the image
of $\L$ in $\pic (\B_X/\B_Y)$, one has an identification
(relative Gauss-Bonnet theorem)
$$c_1([\L])= \frac{i}{2\pi} [K_r] \in \Gamma (Y,DR^2_{\A_X/\A_Y})\,,\tag 4.2$$
where $c_1[\L]\in\Gamma(Y,R^2\pi_\ast\C)$ is regarded as a section in
$\Gamma(Y,R^2\pi_\ast\pi_{-1}\A_Y)\simeq$\break$\Gamma(Y,DR^2_{\A_X/\A_Y})$.

When the family under consideration is a SUSY-curve, it is possible
to state this result in terms
of fiberwise Berezin integration, analogously to the non-graded case.
An added difficulty in this context is that the elements in
$\Gamma(Y,DR^2_{\A_X/\A_Y})$  are not fiberwise integrable,
so that it is necessary to define a new kind of connections, whose
curvature forms are Berezin forms.

\proclaim{Definition} Given a SUSY-curve $(X/Y,\D)$, let $\A^{p,q}$
$(p,q=0,1)$ be the sheaves of Berezin $(p,q)$ forms, and let $\L$ be
a line bundle over $(X,\B_X)$. A conformal connection over $\L$ is
a pair $(\nb,\nbb)$ of even morphisms of graded $\pi^{-1}\B_Y$-modules
$$\nb\colon\L\to \A^{1,0}\otimes_{\B_X}\L,\qquad
\nbb\colon\L\to \A^{0,1}\otimes_{\B_X}\L$$
satisfying the Leibniz rules
$$\nb(f\sigma)=\delta f\otimes\sigma + f\nb(\sigma)\,,\qquad
\nbb(f\sigma)=\bar\delta f\otimes\sigma + f\nbb(\sigma)$$
for all sections $\sigma$ of $\L$ and $f$ of $\B_X$.
\endproclaim

Any (smooth) relative graded connection $\nabla_r$ over $\L$ induces
a conformal connection, by composition with the projections
$\Omega^1_{\A_X/\A_Y}\to\Omega^{1,0}$ or $\Omega^1_{\A_X/\A_Y}\to\Omega^{0,1}$
and with the morphism
$\varrho^{p,q}\colon\Omega^{p,q}\to\A^{p,q}$:
$$\CD
\L @>{\nabla_r}>> \Omega^1_{\A_X/\A_Y}\otimes_{\B_X}\L \\
@. @V{\nb}\bigsearrow\phantom{xxxxxxxxx}VV\\
@. \A^{1,0}\otimes_{\B_X}\L\endCD
\qquad,\qquad
\CD
\L @>{\nabla_r}>> \Omega^1_{\A_X/\A_Y}\otimes_{\B_X}\L \\
@. @V{\nbb}\bigsearrow\phantom{xxxxxxxxx}VV\\
@. \A^{0,1}\otimes_{\B_X}\L
\endCD$$

Given an open cover $\{U_i\}$ of $X$ over which $\L$ trivializes,
and nowhere vanishing local sections $\{\sigma_i\in \L(U_i)\}$,
we define a set of local {\sl conformal connection (Berezin) forms}
$\{\tau_i,\bar\tau'_i\}$ by letting
$$\nb(\sigma_i)=\tau_i\otimes\sigma_i,\qquad
\nbb(\sigma_i)=\bar\tau'_i\otimes\sigma_i.$$
Denoting by $g_{ij}$ the transition functions of $\L$, one proves that
on $U_i\cap U_j$
$$\tau_j=\tau_i+g_{ij}^{-1}\delta g_{ij},\qquad
\bar\tau'_j=\bar\tau'_i+g_{ij}^{-1}\bar\delta g_{ij}\,;$$
these relations characterize the conformal connection, and permit one to show,
by standard partition of unity arguments, the existence of
conformal connections.

One may extend $(\nb,\nbb)$ to a pair of even morphisms
$$\nb\colon\A^{0,1}\otimes_{\B_X}\L\to \A^{1,1}\otimes_{\B_X}\L,\qquad
\nbb\colon\A^{1,0}\otimes_{\B_X}\L\to \A^{1,1}\otimes_{\B_X}\L$$
by letting $\nb(\bar\tau'\otimes\sigma)=\delta(\bar\tau')\otimes\sigma
-\bar\tau'\otimes\nb(\sigma)$ and $\nbb(\tau\otimes\sigma)=
\bar\delta(\tau)\otimes\sigma - \tau\otimes\nbb(\sigma)$.

\proclaim{Definition} The curvature $\hat{K}$ of a conformal connection
$(\nb,\nbb)$ is the morphism
$$\hat{K}=\nb\circ\nbb +\nbb\circ\nb\colon\L\to\L\otimes_{\B_X}\A^{1,1}$$
\endproclaim

\proclaim{Proposition} $\hat{K}$ is a graded $\B_X$-linear morphism.
\endproclaim
\proof By direct computation, using Lemma 3.11.
\qed\enddemo
In view of this result, $\hat{K}$ determines a global section
of $\A^{1,1}$, that we denote by $\hat{K}\in\Gamma(X,\A^{1,1})$ again.
If  the conformal connection $(\nb,\nbb)$
has conformal connection forms $\{\tau_i,\bar\tau'_i\}$,
the local expression of $\hat{K}$ is
$$\hat{K}_i=\bar\delta\tau_i+\delta\bar\tau'_i.$$

If $K_r\in\Gamma (X,\Zc^2_{\A_X/\A_Y})$ is the curvature of
a relative graded connection $\nabla_r$, an easy computation shows that
the curvature $\hat{K}\in\Gamma(X,\A^{1,1})$ of the conformal connection
$(\nb,\nbb)$ induced by $\nabla_r$ is the image of $K_r$ via the morphism
$\Gamma(X,\Zc^2_{\A_X/\A_Y})\to\Gamma(X,\A^{1,1})$ induced by (3.9).

\proclaim{Lemma} The image $[\hat{K}]\in\Gamma(Y,DRB_{X/Y})$ under
the natural morphism \break
$\Gamma(Y,\pi_\ast\A^{1,1})\allowmathbreak\to\allowmathbreak
\Gamma(Y,DRB_{X/Y})$ does not depend on the conformal connection.
\endproclaim
\proof The proof is the usual one: if $(\nb_0,\nbb_0)$,
$(\nb_1,\nbb_1)$ are two conformal connections,
the morphisms
$$\nb_1-\nb_0\colon\L\to \A^{1,0}\otimes_{\B_X}\L\,,\qquad
\nbb_1-\nbb_0\colon\L\to \A^{0,1}\otimes_{\B_X}\L $$
are graded $\B_X$-linear and therefore determine global sections
$\tau\in\Gamma(X,\A^{1,0})\simeq\Gamma(Y,\pi_\ast\A^{1,0})$ and
$\bar\tau'\in\Gamma(X,\A^{0,1})\simeq\Gamma(Y,\pi_\ast\A^{0,1})$;
moreover, the morphisms
$$\nb_t=\nb_0+t\tau\,,\qquad
\nbb_t=\nbb_0+t\bar\tau'$$
define a family of conformal connections, whose curvatures
are $\hat{K}_t =\hat{K}_0 + t (\bar\delta\tau+\delta\bar\tau')$.
Then, the section $\bar\delta\tau+\delta\bar\tau'\in\Gamma(X,\A^{1,1})$
lies in the kernel of $\Gamma(X,\A^{1,1})\to\Gamma(Y,DRB_{X/Y})$,
which proves the claim.
\qed\enddemo

If $[K_r]\in\Gamma(Y,DR^2_{\A_X/\A_Y})$ is the projection of
$K_r$, in view of the commutativity of
diagram (3.10) the image of $[K_r]$ in
$\Gamma(Y,DRB_{X/Y})$ coincides with $[\hat{K}]\in\Gamma(Y,DRB_{X/Y})$,
the image of the curvature $\hat{K}$ of the conformal connection
induced by $\nabla_r$. Then, from (4.2) and Lemma 4.7 one infers that
$$c_1([\L])= \frac{i}{2\pi} [\hat{K}]\in\Gamma(Y,DRB_{X/Y})\,,$$
and by applying the fiberwise Berezin integration one obtains
the following result.
\proclaim {Theorem} {\rm (Gauss-Bonnet)} Let $(X/Y,\D)$ be a SUSY-curve,
$\L$ a line bundle over $(X,\B_X)$, let $[\L]$ be its image in
$\pic (\B_X/\B_Y)$, and let $\hat{K}$ be the curvature of any conformal
connection over $\L$; then
$$\hbox to\hsize{$\phantom{\hbox{\cub}}$\hfill${\dsize
c_1([\L])=\frac{i}{2\pi}\intb \hat{K}}\quad .$\hfill\cub}$$
\endproclaim

\obligedskip\subheading{Acknowledgements} It is a pleasure to
thank C\. Bartocci and D\. Hern\'andez Ruip\'erez for their collaboration
during the early stage of this investigation.
We also acknowledge discussions with
D\. Hern\'andez Ruip\'erez and J\. Mu\~noz Porras about relative duality and
the moduli space of curves.

\obligedskip
\enddocument

\Refs

\ref\key [1] \by Bartocci, C., and Bruzzo, U\. \paper
Cohomology of the structure sheaf of real and complex
supermanifolds \jour J\. Math\. Phys\.
\vol 29 \yr 1988 \pages 1789--1795 \endref

\ref\key [2] \manyby Bartocci, C., Bruzzo, U., and
Hern\'andez Ruip\'erez, D\.
\paper Some results on line bundles over\newline SUSY-curves
\inbook Differential Geometric Methods in Theoretical Physics
\publ Proceedings, Tahoe City 1989,
edited by L.L\. Chau and W\. Nahm, Plenum Press \publaddr
New  York \yr 1991 \pages 667--672 \endref

\ref\key [3] \bysame  \book  The geometry of supermanifolds
\publ Kluwer Acad\. Publ\.
\yr 1991 \publaddr Dordrecht, The Netherlands \endref

\ref\key [4] \by Bernshtein, I.N., and Leites, D.A\.
\paper Integral forms and the Stokes formula on supermanifolds
\jour Funct\. Anal\. Appl\. \vol 11 \yr 1977 \pages 45-47 \endref

\ref\key [5] \by Bruzzo, U., and  Dom\'\i nguez P\'erez, J.A\. \paper
Line bundles over families of (super) Riemann surfaces. I: the non graded case
\jour Preprint Dip\. di Matematica, Univ\. di Genova \yr 1991\endref

\ref\key [6] \by Giddings, S.B., and Nelson, P\.
\paper Line bundles on super Riemann surfaces
\jour Commun\. Math\. Phys\. \vol 118 \yr 1988 \pages 289--302 \endref


\ref\key [7] \by Grothendieck, A\. \paper Sur quelques points
d'alg\`ebre homologique \jour T\^ohoku Math\. J\. \vol 9 \yr 1957 \endref

\ref\key [8] \by Hartshorne, R\. \book Residues and duality \publ
\rm {\sl Lect\. Notes  Math\.} {\bf 20}, Sprin\-ger-Ver\-lag \publaddr
Heidelberg \yr 1966 \endref

\ref\key [9] \by Kostant, B\. \paper Graded manifolds, graded Lie algebras
and prequantization \inbook Differential Geometric Methods in Theoretical
Physics \publ K\. Bleuler and
A. Reetz eds., {\sl Lecture Notes Math.} {\bf 570},
Springer-Verlag \publaddr Berlin \yr 1977 \pages 177--306 \endref


\ref\key [10] \by LeBrun, C., and Rothstein, M\.
\paper Moduli of super Riemann surfaces
\jour Commun\. Math\. Phys\. \vol 117 \yr 1988 \pages 159--176 \endref

\ref\key [11] \by Le\u\i tes, D.A\. (editor) \paper
Seminar on supermanifolds N. 31 \jour Reports of the Department of Mathematics
of the University of Stockholm \vol 14 \yr 1988 \endref

\ref\key [12] \by Levin, A.M\. \paper Supersymmetric elliptic curves
\jour Funct\. Anal\. Appl\. \vol 21 \yr 1987 \pages 243--244 \endref


\ref\key [13] \manyby Manin, Yu. I\. \paper Critical dimensions of the string
theories and the dualizing sheaf of the moduli space of (super) curves
\jour Funct\. Anal\. Appl\. \vol 20 \yr 1986 \pages 244-246 \endref

\ref\key [14] \bysame \book Gauge field theory and complex
geometry \publ {\sl Grund\-leh\-ren der mathematischen
Wissen\-schaften} {\bf 289},
Sprin\-ger-Ver\-lag \publaddr Berlin \yr 1988 \endref

\ref\key [15] \manyby Penkov, I\.B\. \paper $\D$-modules on
supermanifolds \jour Invent\. Math\. \vol 71 \yr
1983 \pages 501--512 \endref


\ref\key [16] \bysame \paper Classical Lie supergroups and Lie
superalgebras and their representations \jour Pr\'e\-pu\-bli\-ca\-tion de
l'Institut Fourier \vol 117 \yr 1988 \endref

\ref\key [17] \manyby Schmitt, T\. \paper Coherent sheaves on analytic
supermanifolds \inbook Seminar Analysis 1983/84
\publ edited by S\. Rempel and B.-W\. Schulze, Akademie der Wissenschaften
der DDR, Institut f\"ur Mathematik
\publaddr Berlin \yr 1984 \pages 94--112 \endref

\ref\key [18] \bysame \paper Some integrability theorems on
supermanifolds \inbook Seminar Analysis 1983/84
\publ edited by S\. Rempel and B.-W\. Schulze, Akademie der Wissenschaften
der DDR, Institut f\"ur Mathematik
\publaddr Berlin \yr 1984 \pages 56--93 \endref

\ref\key [19] \by Vaintrob, A.Yu\. \paper Deformations of complex
superspaces and coherent sheaves on them
\jour J\. Soviet Math\. \vol 51 \yr 1990 \pages 2140--2188 \endref

\ref\key [20] \by Voronov, A.A., Manin, Yu.I., and Penkov I.B\.
\paper Elements of supergeometry
\jour J\. Soviet Math\. \vol 51 \yr 1990 \pages 2069--2083 \endref
\bye